\def\year{2018}\relax
\documentclass[letterpaper]{article} %DO NOT CHANGE THIS
\usepackage{aaai18}  %Required
\usepackage{times}  %Required
\usepackage{helvet}  %Required
\usepackage{courier}  %Required
\usepackage[hyphens]{url}  %Required
\usepackage{graphicx}  %Required
\usepackage{color}
\usepackage{amsmath,amssymb,bm}
\usepackage{booktabs} % For formal tables
\usepackage{balance} % to balance references
\usepackage{subcaption} % subfigure captions
\usepackage{multirow} % for tables with multi-rows and -columns
\usepackage{pbox}
\begin{document}
% The file aaai.sty is the style file for AAAI Press 
% proceedings, working notes, and technical reports.
%
%\title{Multimodal Social Media Crisis Dataset}
%\title{CrisisMMD: Multimodal Crisis-Related Twitter Datasets}
\title{CrisisMMD: Multimodal Twitter Datasets from Natural Disasters}
\author{Firoj Alam, Ferda Ofli, Muhammad Imran\\
Qatar Computing Research Institute, HBKU, Doha, Qatar\\
\{fialam, fofli, mimran\}@hbku.edu.qa
}

\maketitle
\begin{abstract}
During natural and man-made disasters, people use social media platforms such as Twitter to post textual and multimedia content to report updates about injured or dead people, infrastructure damage, and missing or found people among other information types. Studies have revealed that this online information, if processed timely and effectively, is extremely useful for humanitarian organizations to gain situational awareness and plan relief operations. In addition to the analysis of textual content, recent studies have shown that imagery content on social media can boost disaster response significantly. Despite extensive research that mainly focuses on textual content to extract useful information, limited work has focused on the use of imagery content or the combination of both content types. One of the reasons is the lack of labeled imagery data in this domain. Therefore, in this paper, we aim to tackle this limitation by releasing a large multimodal dataset collected from Twitter during different natural disasters. We provide three types of annotations, which are useful to address a number of crisis response and management tasks for different humanitarian organizations.\vspace{-2mm}

%Current literature shows significant amount of work on the use of textual content towards developing automated system to provide useful information. Very recently, a few studies explored the use of imaginary content and its usefulness. Hence, current state-of-art is limited either text or image only human labeled crisis related dataset. Therefore, in this study we aim to address this limitation. We have collected tweets during eight natural disasters focusing on four disaster types. Then, manually annotated them using crowd-sourcing for three different tasks such as informative \textit{vs.} not-informative, humanitarian categories and damage severity.  

\end{abstract}

%%%
% keywords: Multimodal, Twitter datasets, Textual and multimedia content, Natural disasters
%%%

% ================
% Introduction
% ================
\section{Introduction}
\label{sec:introduction}
At times of natural and man-made disasters, social media platforms such as Twitter and Facebook are considered vital information sources that contain a variety of useful information such as reports of injured or dead people, infrastructure and utility damage, urgent needs of affected people, and missing or found people among others~\cite{houston2015social,Alam2018}. Information shared on social media has a wide variety of applications~\cite{imran2014aidr,ashktorab2014tweedr,reuter2015xhelp,poblet2014crowdsourcing,Kishi2017,LaudyClaire2017,Meissen2017}. One application that also motivates our work is ``humanitarian aid" where the primary purpose of humanitarian organizations such as The United Nations Office for the Coordination of Humanitarian Affairs (OCHA) is to gain situational awareness and actionable information to save lives, reduce the suffering of affected people, and rebuild communities~\cite{CarlosCastillo2016}.

Processing social media data to extract life-saving information which is also helpful for humanitarian organizations in preparedness, response, and recovery of an emergency involves solving multiple challenges including handling information overload, information classification and determining its credibility, prioritizing certain types of information, etc.~\cite{imran2015processing}. These challenges require building computational systems and methods useful for a number of information processing tasks such as information classification, clustering, and summarization among others.   

Information on social media is mainly shared in two forms: textual messages and images. Most of the past studies and systems mainly focused on using textual content to aid disaster response. However, in addition to the usefulness of textual messages, recent studies have revealed that images shared on social media during a disaster event can help humanitarian organizations in a number of ways. For example, \citeauthor{nguyen17damage} used images shared on Twitter to assess the severity of infrastructure damage~\cite{nguyen17damage}. 
%\textcolor{red}{XYZ studies images from X event and found their utility to ABC humanitarian tasks.}
\citeauthor{petersinvestigating} reported that the existence of images within on-topic messages were more relevant to the disaster event based on their analysis of tweets and messages from Flickr and Instagram for the flood event in Saxony in 2013~\cite{petersinvestigating}. Similarly, \citeauthor{jing2016integration} investigated the usefulness of image and text and found that they were both informative. For their study, they collected data from two sources related to flood and flood aid \cite{jing2016integration}. A similar study has been conducted by \cite{kelly2017mining} to extract useful information from flood events occurred in Ireland during December 2015 to January 2016. 

Despite extensive research that mainly focuses on social media text messages, limited work has focused on the use of images to boost humanitarian aid. One reason that hinders the growth of this research line is the lack of ground-truth data. There exist a few repositories such as CrisisLex~\cite{olteanu2014crisislex} and CrisisNLP~\cite{imran2016lrec} %such as \textcolor{red}{only a few? I think there are plenty...} 
which offer several Twitter datasets from natural and man-made disasters, %, \textcolor{red}{also try to include some pointers to datasets other than ours only} 
but all of them share only textual content annotations. To overcome this limitation, we present human-labeled multimodal datasets collected from Twitter during seven recent natural disasters including earthquakes, hurricanes, wildfires, and floods. To the best of our knowledge, these are the first multimodal Twitter datasets ever shared publicly with ground-truth annotations.\footnote{The dataset is available at \url{https://dataverse.mpi-sws.org/dataverse/icwsm18}}

To acquire ground-truth labels, we employed paid workers from a well-known crowdsourcing platform (i.e., Figure Eight\footnote{\url{https://www.figure-eight.com/}, previously known as CrowdFlower, \url{http://crowdflower.com/}}) and asked them to annotate data based on three humanitarian tasks. The first task aims to determine the informativeness of a given tweet text or an image for humanitarian aid purposes. Given the fact that millions of tweets are shared during disasters, focusing only on the \emph{informative} messages or images help reduce information overload for humanitarian organizations. The second task aims to further analyze the set of messages and images that have been identified as informative in the first task to determine what kind of humanitarian information they convey (see Section \textit{Humanitarian Tasks and Manual Annotations} for detailed categories). Finally, the third task aims to assess the severity of damage to infrastructure and utilities observed in an image.
%We will share the datasets online upon the acceptance of the paper.
%Furthermore, to show the utility of the annotated datasets, we conduct experiments. 

The rest of the paper is organized as follows. In the next section, we provide a summary of the related work. Then, we provide details about the disaster events and the data collection procedure in our study. Next, we elaborate on the humanitarian tasks as well as their annotation details and results. 
%After that, we present our baseline experiments. 
Furthermore, we present possible applications and discussion in the later section. Finally, we conclude the paper in the last section.

% ================
% Related Work
% ================
\section{Related Work}
\label{sec:related_work}
The use of social media such as Twitter, Facebook, and Youtube, has been explored in numerous studies \cite{imran2014aidr,vieweg2010microblogging,imran2015processing,terpstra2012towards,tsou2017building} for curating, analyzing and summarizing crisis-related information in order to make some decisions and responses. Current literature does not only highlight its importance but also provides directions for possible research avenues. Among them, one of the important research avenues is the exploitation of textual and visual content to extract useful information for humanitarian aid, which has been remained unexplored to a large extent. One of the important limitations of this line of research is the lack of ground-truth data. Below, we describe works that provide crisis-related datasets.

In crisis informatics, one of the earliest and publicly-available datasets is CrisisLex~\cite{olteanu2014crisislex}. It consists of tweets collected during six disaster events occurred in USA, Australia, and Canada between October 2012 and July 2013. The dataset was collected using keywords and geo-graphical information from Twitter. The annotations of the dataset consist of i) directly related, ii) indirectly related, iii) not related, and iv) not in English or not understandable.
%The dataset provides annotations along the relatedness of a tweet to the disaster event. 
%\textcolor{red}{I did not fully understand the previous sentence.}
%They annotated data using Crowdflower and for each event they sampled ~10K tweets for the annotation. The task of the annotator was to assign a label with either a) directly related to disaster, b) indirectly related, c) not related or d) not related or no in English. 
%
In another work~\cite{olteanu2015expect}, the authors provide a dataset that consists of tweets from 26 crisis events that took place between 2012 and 2013. In this work, they first characterize the datasets along different crisis dimensions: 1) hazard type (i.e., natural vs.\ human-induced) and their subtypes, 2) temporal development (i.e., instantaneous vs.\ progressive), 3) geographic (i.e., focalized vs.\ diffused). Then, they characterized the datasets by 1) informativeness, 2) information type, and 3) source. Similar to the previous study they employed crowd-source workers to annotate the dataset. The dataset is publicly available at CrisisLex site\footnote{\url{http://crisislex.org}}.

Another initiative to provide crisis-related data is CrisisNLP\footnote{\url{http://crisisnlp.qcri.org}}. Currently, this site has published three major data resources. For instance, \citeauthor{imran2013practical} provide tweets collected during the Joplin tornado, which hit Joplin, Missouri (USA) on May 22, 2011, and the tweets collected during the Hurricane Sandy, which hit Northeastern US on October 29, 2012~\cite{imran2013practical}. The annotated dataset consists of 2,000 tweets for Hurricane Sandy and about 4,400 for Joplin Tornado. Recently published dataset by \cite{imran2016lrec} consists of tweets from 19 different crisis events that took place between 2013 to 2015. A particular focus of this dataset is humanitarian information categories annotated by Stand-By-Task-Force (STBF) volunteers and crowd-workers from CrowdFlower. In another study, \citeauthor{ashktorab2014tweedr} report a dataset that has been collected from 12 different crises occurred in the United States~\cite{ashktorab2014tweedr}. The annotation of this dataset consists of infrastructure damage and human casualty.
%In addition, it also consists of annotation type of the infrastructure damage.
% \cite{castillo2016big}
%\cite{temnikova2015emterms}
%A recent study \cite{nguyen17damage} presents a damage assessment image dataset collected from the different crisis event. The annotation consists of damage severity levels such as 1) severe damage, 2) mild damage, 3) little-to-no damage with around $\sim27K$ images.
\citeauthor{wang2015hurricane} present a corpus of 6.5 million geotagged tweets collected during 2012 Hurricane Sandy~\cite{wang2015hurricane}. However, this corpus does not provide any human labeled annotations. \citeauthor{lagerstrom2016image} present the utility of image classification to support emergency situation by utilizing tweets collected from the event of 2013 New South Wales bushfires~\cite{lagerstrom2016image}. They have ${\sim}$5,000 images with labels ``fire'' or ``not-fire'' and present an image classification accuracy of 86\%. The limitation is that their data is not publicly available for research.

Despite all the initiatives for providing crisis-related datasets that are mainly useful for natural language processing tasks, no multimodal dataset consisting of combined textual and visual annotations has been published yet.
%Upon looking into these datasets, it is clear that they are either text or image only dataset. Hence, the current state of art lacks a dataset which combines text and image for humanitarian tasks. 
In this paper, we try to bridge this gap by releasing multimodal datasets that are collected from Twitter during seven natural disasters in 2017 and annotated for several tasks. We hope the research community will take advantage of this multimodal dataset to advance the research on both image and text processing.
%This study address this limitation by collecting, annotating and making them publicly available.

% ================
% Data Collection
% ================
\section{Natural Disaster Events and Data Collection}
\label{sec:data_collection}
We used Twitter to collect data during seven natural disasters. The data collection was performed using event-specific keywords and hashtags. In Table~\ref{table:events_keywords}, we list the keywords used and the data collection period for each event. Next, we provide details of data collection for each event. %In table~\ref{table:collected_data_distribution}, we show detailed information about the collected data, filtered and sampled items. %present the keywords we used, start and end date, and some other statistics. 

% \begin{table*}[h]
% \centering
% \caption{CrisisMMD dataset details: type, name, keywords use and their start and end date. All events occurred in 2007.}
% \label{table:events_keywords}
% \begin{tabular}{@{}lllll@{}}
% \toprule
% \textbf{Crisis type} & \textbf{Crisis Name} & \textbf{Keywords} & \textbf{Start date} & \textbf{End date} \\ \midrule
% \textbf{Wilefire} & \textbf{California Wilefire} & \pbox{7cm}{California fire, California wildfire, Wildfire California, USA Wildfire, California wildfires} & Oct 10 & Oct 27 \\
% \textbf{Flood} & \textbf{Srilanka Flood} & \pbox{7cm}{flood Sri Lanka, FloodSL, SriLanka flooding, SriLanka floods, SriLanka flood, typhoon mora, cyclone mora, mora, CycloneMora} & May 31& Jul 3 \\
% \textbf{Hurricane} & \textbf{Harvey} & \pbox{7cm}{Hurricane Harvey, Harvey, HurricaneHarvey, Tornado} & Aug 26 & Sep 20\\
% \textbf{Hurricane} & \textbf{Irma} & \pbox{7cm}{Hurricane Irma, Irma storm, Storm Irma, Irma Hurricane, Irma} & Sep 6& Sep 21\\
% \textbf{Hurricane} & \textbf{Maria} & \pbox{7cm}{Hurricane Maria, Maria Storm, Maria Cyclone, Maria Tornado, Tropical Storm Maria, HurricaneMaria, puerto rico} & Sep 20& Nov 13\\
% \textbf{Earthquake} & \textbf{Iraq Earthquake} & \pbox{7cm}{kuwait earthquake, iran earthquake, halabja earthquake, Iraq earthquake} & Nov 13& Nov 19\\
% \textbf{Earthquake} & \textbf{Mexico Earthquake} & \pbox{7cm}{mexico earthquake, mexicoearthquake} & Sep 20 & Oct 6\\ \bottomrule
% \end{tabular}
% \end{table*}

\begin{table*}[h]
	\centering
	\caption{CrisisMMD dataset details including event names, keywords used for data collection, and data collection period.}
	\label{table:events_keywords}
	\begin{tabular}{@{}llrr@{}}
		\toprule
		\textbf{Crisis event} & \textbf{Keywords} & \textbf{Start date} & \textbf{End date} \\ \midrule
		
		\textbf{Hurricane Irma} & \pbox{10cm}{\it Hurricane Irma, HurricaneIram, Irma storm,...} & Sep 6, 2017& Sep 21, 2017\\
		
		\textbf{Hurricane Harvey} & \pbox{10cm}{\it Hurricane Harvey, Harvey, HurricaneHarvey,...} & Aug 26, 2017 & Sep 20, 2017\\
		
		\textbf{Hurricane Maria} & \pbox{10cm}{\it Hurricane Maria, Maria Storm, Maria Cyclone,...} & Sep 20, 2017 & Nov 13, 2017\\
		
		\textbf{Mexico earthquake} & \pbox{10cm}{\it Mexico earthquake, mexicoearthquake,...} & Sep 20, 2017 & Oct 6, 2017\\ 
		
		\textbf{California wildfires} & \pbox{10cm}{\it California fire, California wildfires,...} & Oct 10, 2017 & Oct 27, 2017 \\
		
		\textbf{Iraq-Iran earthquake} & \pbox{10cm}{\it Iran earthquake, Iraq earthquake, halabja earthquake,...} & Nov 13, 2017 & Nov 19, 2017\\
		
		\textbf{Sri Lanka floods} & \pbox{10cm}{\it SriLanka floods, FloodSL, SriLanka flooding,...} & May 31, 2017 & Jul 3, 2017 \\
		\bottomrule
	\end{tabular}
\end{table*}

\subsection{Hurricane Irma 2017}
\label{ssec:hurricane_irma}
Hurricane Irma\footnote{\url{https://en.wikipedia.org/wiki/Hurricane_Irma}} caused catastrophic damage in Barbuda, Saint Barthelemy, Saint Martin, Anguilla, and the Virgin Islands. On Friday, September 8, a hurricane warning was issued for the Florida Keys and the Florida governor ordered all public schools and colleges to be closed. The Irma storm was a Category 5 hurricane, which caused $\$66.77$ billion in damage. We collected Hurricane Irma-related data from Twitter starting from September 6, 2017, to September 19, 2017, and the resulted collection consists of ${\sim}3.5$ million tweets and ${\sim}$176,000 images.
%using the keywords: \textit{``Hurricane Irma'', ``Irma storm'', ``Irma''}. In total, 1,189,384 tweets were collected during this period.
% Figure~\ref{fig:daily_tweet_volume} (middle chart) shows the distribution of daily tweets of Hurricane Irma data. On the first day (i.e., September 6), we can see a surge of tweets in which more than 300,000 tweets were collected. However, during the next days the distribution stayed steady around 50,000 tweets per day. Besides, 60,932 of these Hurricane Irma tweet data contained unique image URLs.

\subsection{Hurricane Harvey 2017}
\label{ssec:hurricane_harvey}
Hurricane Harvey was a Category 4 storm when it hit Texas, USA on August 25, 2017\footnote{\url{https://en.wikipedia.org/wiki/Hurricane_Harvey}}. It caused nearly $\$200$ billion in damage, which is record-breaking compared with any natural disaster in the US history. As can be seen in Table~\ref{table:events_keywords}, we started the data collection on August 25, 2017, and ended on September 5, 2017. In total, ${\sim}7$ million tweets with ${\sim}$300,000 images were collected during this period.
%Among these Hurricane Harvey collected data, $\sim300K$ tweets were found having one or more images. 
%Many images seem duplicate, so we performed image URL-based de-duplication, which resultantly left ~111,725 images with unique image URLs.

\subsection{Hurricane Maria 2017}
\label{ssec:hurricane_maria}
Hurricane Maria\footnote{\url{https://en.wikipedia.org/wiki/Hurricane_Maria}}, was a Category 5 hurricane that slammed Dominica and Puerto Rico and caused more than 78 deaths including 30 in Dominica and 34 in Puerto Rico, while many more left without homes, electricity, food, and drinking water. The data collection for Hurricane Maria was started on September 20, 2017, and ended on October 3, 2017. In total, we collected ${\sim}3$ million tweets and ${\sim}$52,000 images. 
%The right chart in Figure~\ref{fig:daily_tweet_volume} shows the daily tweet distribution for the Hurricane Maria data. Of these Hurricane Maria tweet data, we found 19,681 tweets with unique image URLs.

\subsection{California Wildfires 2017}
\label{ssec:california_wildfire}
A series of wildfire took place in California in October 2017\footnote{\url{https://en.wikipedia.org/wiki/2017_California_wildfires}} causing more than $\$9.4$ billion losses of property. We started our tweet collection on October 10, 2017 and continued until October 27, 2017. As can be seen in Table~\ref{table:collected_data_distribution}, the collected dataset contains ${\sim}$400,000 tweets and ${\sim}$10,000 images. 

\subsection{Mexico Earthquake 2017}
\label{ssec:mexico_earthquake}
The Mexico earthquake\footnote{\url{https://en.wikipedia.org/wiki/2017_Central_Mexico_earthquake}} on September 19, 2017 was another major earthquake with a magnitude of $7.1$. The earthquake caused death of around 370 people. For this event, our data collection started on September 20, 2017 till October 6, 2017. In total, we collected ${\sim}$400,000 tweets and ${\sim}$7,000 images.   

\subsection{Iraq-Iran Border Earthquake 2017}
\label{ssec:iraq_earthquake}
On November 12, 2017, a strong earthquake with a magnitude of $7.3$ struck the border of Iran and Iraq\footnote{\url{https://en.wikipedia.org/wiki/2017_Iran–Iraq_earthquake}}.
%The Iraq earthquake on November 12, 2017,  is one of the strongest earthquake recorded in the region since 1967 with a magnitude of 6.1. 
The earthquake caused around 630 casualties, seventy thousand became homeless and eight thousand were injured. For our study, we collected tweets from November 12, 2017 to November 19, 2017, which resulted in ${\sim}$200,000 tweets and ${\sim}$6,000 images. 

\subsection{Sri Lanka Floods 2017}
\label{ssec:srilanka_flood}
Due to heavy monsoon on southwest, Sri Lanka faced severe flooding in May 2017\footnote{\url{https://en.wikipedia.org/wiki/2017_Sri_Lanka_floods}}. %The Sri Lanka floods that took place in May, 2017, which was resulted from a heavy southwest monsoon. 
Furthermore, the flooding situation got worsened due to the Cyclone Mora\footnote{\url{https://en.wikipedia.org/wiki/Cyclone_Mora}}, which caused more floods and landslides throughout Sri Lanka during the last week of May 2017. Our tweet data collection started on May 31, 2017 until July 3, 2017, which resulted in ${\sim}$41,000 tweets and ${\sim}$2,000 images.  %For collecting the tweets we started the collection on May 31 and collected them until July 3rd, 2017, which resulted in $\sim41K$ tweets and $\sim2K$ images.  

\begin{table*}[h]
	\centering
	\caption{Event-wise data distribution. The numbers inside the parentheses in the last column represent the total number of images associated with tweets. Total number of images can be larger than total number of tweets as some tweets contain more than one image.}
	\label{table:collected_data_distribution}
	\scalebox{1.0}{
		\begin{tabular}{@{}lrrrr@{}}
			\toprule
			% \multicolumn{1}{c}{\textbf{Crisis name}} & \multicolumn{1}{c}{\textbf{\# tweets}} & \multicolumn{1}{c}{\textbf{\# images}} & \multicolumn{1}{c}{\textbf{\# filtered tweets}} & \multicolumn{1}{r}{\textbf{\# sampled tweets (images)}} \\ \midrule
			\textbf{Crisis name} & \textbf{\# tweets} & \textbf{\# images} & \textbf{\# filtered tweets} & \textbf{\# sampled tweets (images)} \\ \midrule
			
			\textbf{Hurricane Irma} &  3,517,280 & 176,972 & 5,739 & 4,041 (4,525) \\
			\textbf{Hurricane Harvey} &  6,664,349 & 321,435 & 19,967 & 4,000 (4,443) \\
			\textbf{Hurricane Maria} &  2,953,322 & 52,231 & 6,597 & 4,000 (4,562) \\
			\textbf{California wildfires} &  455,311 & 10,130 & 1,488 & 1,486 (1,589) \\
			\textbf{Mexico earthquake} &  383,341 & 7,111 & 1,241 & 1,239 (1,382) \\
			\textbf{Iraq-Iran earthquake} &  207,729 & 6,307 & 501 & 499 (600) \\
			\textbf{Sri Lanka floods} &  41,809 & 2,108 & 870 & 832 (1,025) \\
			\midrule
			\textbf{Total}  & {\bf 14,223,141} & {\bf 576,294} & {\bf 36,403} &  {\bf 16,097 (18,126)}\\ 
			\bottomrule
		\end{tabular}
	}
\end{table*}

\subsection{Data Filtering and Sampling}
\label{ssec:preprocessing}
To prepare data for manual annotation, we perform the following filtering steps: %The preprocessing part includes the following steps:
\begin{enumerate}
	\item As we build a multimodal dataset, we are interested only in tweets \emph{with} images. Thus, our first filtering step is to discard all the tweets that do not contain at least one image URL. Since a tweet can contain more than one image, we extract all image URLs from the {\it ``extended\_entities''} element of the retrieved JSON record of a tweet. %next we  Filtering out tweets containing no image URLs. We check extended entities in the json file that Twitter API returns. The reason for checking extended entities is that it stores more than one image URLs if a user posts more than one images. An important fact is that there are many duplicate image URLs and we filter them out before start crawling the images.
	\item We discard all non-English tweets using Twitter-provided language meta-data for a given tweet. 
	\item We retain tweets that contain at least two or more words or hashtags. In other words, we remove tweets containing a single word or hashtag since single-word tweets are less likely to convey any useful or meaningful information. We do not consider URLs or numbers as proper English words.%Filtering out non-English tweets, removing URLs, numbers and special characters.
	%\item Remove tweets containing one word.
	\item We remove duplicate tweets using tweets' textual content. For this purpose, we use the cosine similarity measure to compute tweet similarity scores. Two tweets with a similarity score greater than $0.7$ are considered duplicate. %compute a vector using bi-grams (i.e., two words) and  Remove duplicate tweets. To check the duplicate we compute a vector from character bigrams and then compute the cosine similarity. We consider a tweet as duplicate compared to an existing tweet if its similarity measure is greater than a threshold of $0.7$. To keep track of existing tweets we maintain an in-memory list. 
\end{enumerate}
After performing the above mentioned filtering steps, we take a random sample of $N$ tweets containing one or more images from each dataset. Due to budget limitations, we sample around 4,000 for Hurricanes Irma, Harvey, and Maria. For the rest, we take all of the filtered tweets, as they are already low numbers. Table~\ref{table:collected_data_distribution} describes all the datasets with details including total number of tweets initially collected, total number of images associated with the initial set of tweets, and the total number of tweets retained after the filtering and sampling steps for each dataset. In particular, the last column of the table shows the number of tweets and corresponding images (in parentheses) for each disaster event in our dataset. A tweet can contain more than one image, and hence, the number of images (shown in parentheses) are slightly larger than the actual number of sampled tweets.

%As shown in Table \ref{table:collected_data_distribution}, in different collections 2\% to 5\% tweets contain images. In column 4th we present a number of filtered tweets, which consist of $\sim38K$ tweets in total for seven collections. Annotating such a high number of tweets for three task is costly. Hence, we sampled tweets for the annotation tasks. In the last column of Table \ref{table:collected_data_distribution}, the number inside the parenthesis represents the number of images associated with tweets. As discussed earlier for a tweet, we crawled one or more images stored in extended entities, hence, we see that there are more images than tweets' text and we used them all for the annotation tasks.  

% \begin{figure}
% \centering
% \includegraphics[width=1\linewidth]{figs/929991089150406656_0.jpg}
% \caption{IFTTT image}
% \label{fig:info_tweets_dist}
% \end{figure}

%In the next Section, we discuss the annotation task and annotation process using the crowdflower platform.

% ================
% Data Collection
% ================

%\input{sec_annotation}
\section{Humanitarian Tasks and Manual Annotations}
\label{sec:annotation}
We perform the manual annotations of the sampled data along three humanitarian tasks. The first task aims to categorize the data into two high-level categories called ``Informative" or ``Not informative". During disasters and emergencies, as thousands of tweets arrive per minute, determining whether or not a tweet contains crucial information useful for humanitarian aid is an important task to reduce information overload for humanitarian organizations. 

The second task, on the other hand, aims to identify critical and potentially actionable information such as reports of injured or dead people, infrastructure damage, etc. from the tweets. For this purpose, we use seven humanitarian categories.
%, namely, ``Infrastructure and utility damage'', ``vehicle damage'', ``Rescue, volunteering, or donation effort'', ``Injured or dead people'', ``Affected individuals'', ``Missing or found people'', ``Other relevant information'', and ``Not relevant or can't judge''.
The third task is specific to damage severity assessment from images. Determining severely-damaged critical infrastructure after a major disaster is a core task of many humanitarian organizations to direct their response efforts.

Next, we present the exact instructions provided to the human annotators for all three tasks.

\subsection{Task 1: Informative vs.\ Not informative}
\label{ssec:info_no_info}
%\subsubsection{Task description:} 
The purpose of this task is to determine whether a given tweet or image, which was collected during {\it``event name''}, is useful for humanitarian aid purposes as defined below. If the given tweet/image is useful for humanitarian aid, it is considered as an ``Informative" tweet/image, otherwise as a ``Not informative" tweet/image.

\subsubsection{``Humanitarian aid" definition:} 
In response to humanitarian crises including natural and man-made disasters, humanitarian aid involves providing assistance to people who need help. The primary purpose of humanitarian aid is to save lives, reduce suffering, and rebuild affected communities. Among the people in need belong homeless, refugees, and victims of natural disasters, wars, and conflicts who need basic necessities like food, water, shelter, medical assistance, and damage-free critical infrastructure and utilities such as roads, bridges, power-lines, and communication poles. 

Moreover, the tweet/image is considered ``Informative'' if it reports/shows one or more of the following: cautions, advice, and warnings, injured, dead, or affected people, rescue, volunteering, or donation request or effort, damaged houses, damaged roads, damaged buildings; flooded houses, flooded streets; blocked roads, blocked bridges, blocked pathways; any built structure affected by earthquake, fire, heavy rain, strong winds, gust, etc., disaster area maps.

Images showing banners, logos, and cartoons are \emph{not} considered as ``Informative''.

\begin{itemize}
	\item \textbf{Informative:} if the tweet/image is useful for humanitarian aid.
	\item \textbf{Not informative:} if the tweet/image is not useful for humanitarian aid.
	\item \textbf{Don't know or can't judge:} due to non-English tweet or low-quality image content.
\end{itemize}

%\subsubsection{Example:}
%TODO

\subsection{Task 2: Humanitarian Categories}
%\subsubsection{Task description:} 
The purpose of this task is to understand the type of information shared in an image/tweet, 
% The below data
which was collected from Twitter during {\it``event name''}. Given an image/tweet, categorize it into one of the following categories.

%Categories list:
\begin{itemize}
	
	\item \textbf{Infrastructure and utility damage:} if the tweet/image reports/shows any built structure affected or damaged by earthquake, fire, heavy rain, floods, strong winds, gusts, etc. such as damaged houses, roads, buildings; flooded houses, streets, highways; blocked roads, bridges, pathways; collapsed bridges, power lines, communication poles, etc.
	
	\item \textbf{Vehicle damage:} if the tweet/image reports/shows any type of damaged vehicle such as cars, trucks, buses, motorcycles, boats, ships, trams, trains, etc.
	
	\item \textbf{Rescue, volunteering, or donation effort:} if the tweet/image reports/shows any type of rescue, volunteering, or donation effort such as people being transported to safe places, people being evacuated from the hazardous area, people receiving medical aid or food, people in shelter facilities, donation of money, blood, or services etc.
	
	\item \textbf{Injured or dead people:} if the tweet/image reports/shows injured or dead people.
	
	\item \textbf{Affected individuals:} if the tweet/image reports/shows people affected by the disaster event such as people sitting outside; people standing in queues to receive aid; people in need of shelter facilities, etc.
	
	\item \textbf{Missing or found people:} if the tweet/image reports/shows instances/pictures of missing or found people due to the disaster event.
	
	\item \textbf{Other relevant information:} if the tweet/image does not belong to any of the above categories, but it still contains important information useful for humanitarian aid, then select this category.
	
	\item \textbf{Not relevant or can't judge:} if the image is irrelevant or you can't judge, for example, due to its low-quality.
	
\end{itemize}

%\subsubsection{Example:}

\subsection{Task 3: Damage Severity Assessment}
The purpose of this task is to assess the severity of damage reported/shown in an image. The severity of damage is the extent of physical destruction to a build-structure. We are only interested in physical damages like broken bridges, collapsed or shattered buildings, destroyed or cracked roads, etc. An example of a non-physical damage is the sign of smoke. Damage severity categories are discussed below:
\begin{itemize}
	\item \textbf{Severe damage:} Substantial destruction of an infrastructure belongs to the severe damage category. For example, a non-livable or non-usable building, a non-crossable bridge, or a non-driveable road are all examples of severely damaged infrastructures.
	
	Specifically,
	\begin{itemize}
		\item \textbf{Building:} If one or more buildings in the focus of the image show substantial loss of amenity/roof. If the image shows a building that is unsafe to use, it should be marked as severe damage.
		
		\item \textbf{Bridge:} If a bridge is visibly not safe to use because parts of it are collapsing and should not be driven or walked upon, it should be listed as severe damage.
		
		\item \textbf{Road:} If a road should not be used because there has been substantial damage, it should be marked as severe damage. Examples: due to an avalanche, there may be huge rocks piled up and you cannot drive or only a narrow part of the road is open.  Due to an earthquake, you see a sinkhole, a substantial part of the road has sunk and the road cannot be navigated safely, that is severe damage.  
	\end{itemize}
	
	\item \textbf{Mild damage:} Partially destroyed buildings, bridges, houses, roads belong to mild damage category. 
	
	\begin{itemize}
		\item \textbf{Building:} Damage generally exceeding minor [damage] with up to 50\% of buildings in the focus of the image sustaining a partial loss of amenity/roof.  Maybe only part of the building has to be closed down, but other parts can still be used.
		
		\item \textbf{Bridge:} If the bridge can still be used, but, part of it is unusable and/or needs some amount of repairs. 
		
		\item \textbf{Road:}  If the road is still usable, but part of it has to be blocked off because of damage.  This damage should be substantially more than what we see due to regular wear or tear.
		
	\end{itemize}
	\item \textbf{Little or no damage:} Images that show damage-free infrastructure (except for wear and tear due to age or disrepair) belong to the little-or-no-damage category.
	
	\item  \textbf{Don't know or can't judge:} Due to low-quality image.
	
\end{itemize}

%\subsubsection{Example:}

\subsection{Manual Annotations using Crowdsourcing}
Given the above specified tasks and instructions, we used Figure Eight, which is a well-known paid crowdsourcing platform previously known as CrowdFlower, 
%\footnote{\url{https://www.figure-eight.com/} or \url{http://crowdflower.com/}}
to acquire manual annotations of the sampled data. Manual annotations for tweets (textual content) and images were acquired separately. In this case, a task consisted of an image or a tweet along with task instructions and a list of categories (e.g., informative and not informative). We first ran Task 1 (i.e., informative or not informative) for all the events. For Task 2 (i.e., humanitarian categories), we only used the data which was labeled as ``informative" (i.e., either text or image was informative) in Task 1. We dropped tweets where neither text nor image was informative. For Task 3 (i.e., damage assessment), we only used images from Task 2 which were labeled as ``Infrastructure and utility damage". For each task, we created at least 40 test questions to keep good quality annotators while excluding annotators that do not perform well on the test questions. We sought an agreement of three different human annotators to decide a final label/category for a tweet or an image. Human annotators with English language expertise were allowed to perform the tasks.

% \begin{figure*}[h]
\begin{figure*}[htbp!]
	\renewcommand{\arraystretch}{0.5} % this reduces the vertical spacing between rows
	\linespread{0.5}\selectfont\centering
	\resizebox{.85\linewidth}{!}{%
		\begin{tabular}{p{0.1cm} p{0.28\textwidth} p{0.28\textwidth} p{0.28\textwidth}}
			%\begin{tabular}{l c c c}
			&
			\textbf{Hurricane Maria}
			&
			\textbf{California Wildfires}
			&
			\textbf{Mexico Earthquake}
			\\
			\raisebox{4.5\normalbaselineskip}[0pt][0pt]{\rotatebox[origin=c]{90}{\textbf{Informative}}}
			&
			\includegraphics[width=0.28\textwidth]{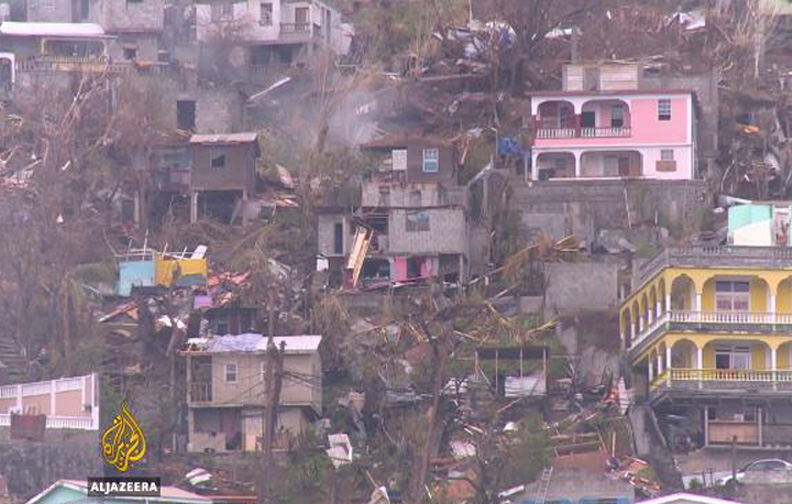}
			&
			\includegraphics[width=0.28\textwidth]{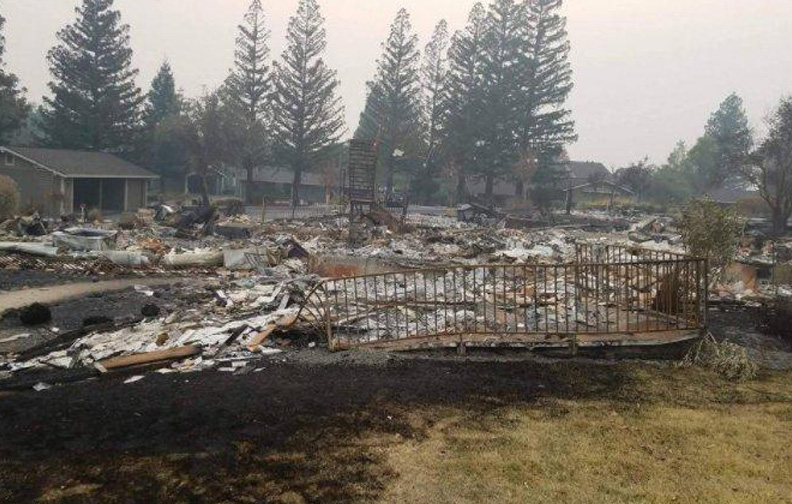}
			&
			\includegraphics[width=0.28\textwidth]{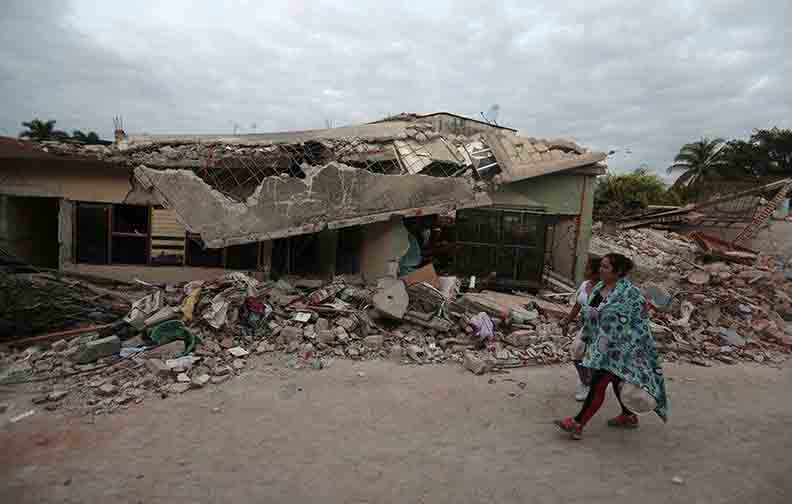}
			\\
			&
			{\scriptsize \textbf{(a)} Hurricane Maria turns Dominica into 'giant debris field' https://t.co/rAISiAhMUy by \#AJEnglish via {@}c0nvey https://t.co/I4zeuW4gkc}
			&
			{\scriptsize \textbf{(b)} A friend's text message saved Sarasota man from deadly California wildfire https://t.co/0TNMFgL885 https://t.co/CIzo44Npza}
			&
			{\scriptsize \textbf{(c)} Earthquake leaves hundreds dead, crews combing through rubble in \#Mexico https://t.co/XPbAEIBcKw https://t.co/wGVxGD4xNd}
			\\
			\raisebox{5.8\normalbaselineskip}[0pt][0pt]{\rotatebox[origin=c]{90}{\textbf{Not informative}}}
			&
			\includegraphics[width=0.28\textwidth]{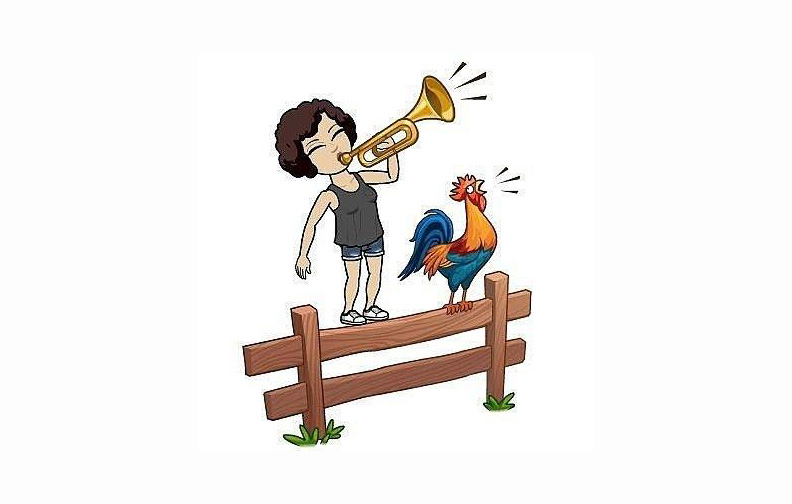}
			&
			\includegraphics[width=0.28\textwidth]{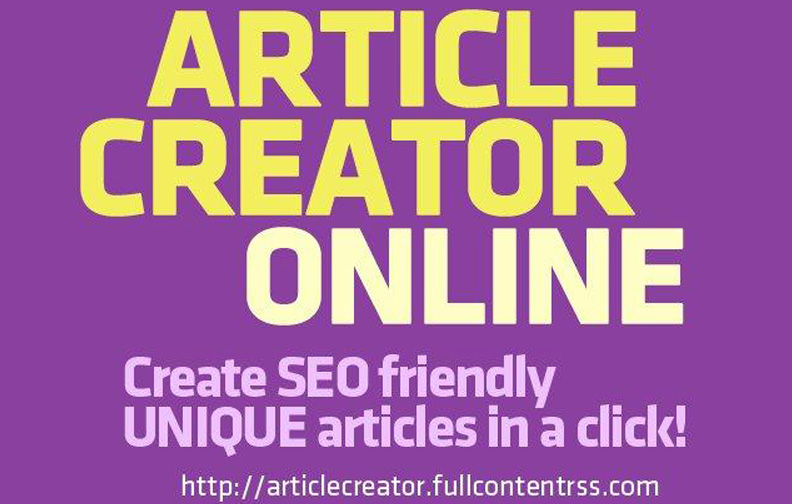}
			&
			\includegraphics[width=0.28\textwidth]{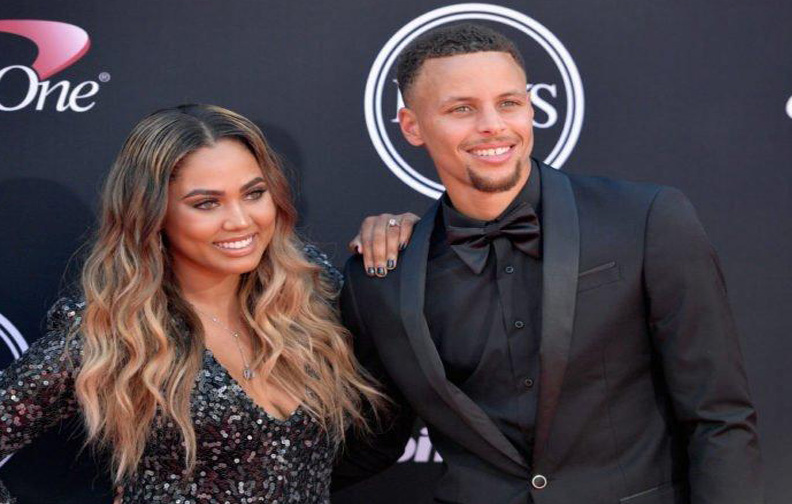}
			\\
			&
			{\scriptsize \textbf{(d)} @SueAikens hi su o back againe big hug FROM PUERTO RICO love you https://t.co/HCEyIHB0QZ}
			&
			{\scriptsize \textbf{(e)} https://t.co/jh0aQql3dR SEO ARTICLE GENERATOR https://t.co/2108RuhxgY \#blogging \#backlinks | Nurse fleeing California wildfires}
			&
			{\scriptsize \textbf{(f)} SEASON OVER???? WE COULD USE ABLE BODIES AT EARTHQUAKE IN MEXICO! DIG IN.... https://t.co/QlnYHtv9AI}
			\\
			\raisebox{5.2\normalbaselineskip}[0pt][0pt]{\rotatebox[origin=c]{90}{\textbf{Rescue \& volunteering}}}
			&
			\includegraphics[width=0.28\textwidth]{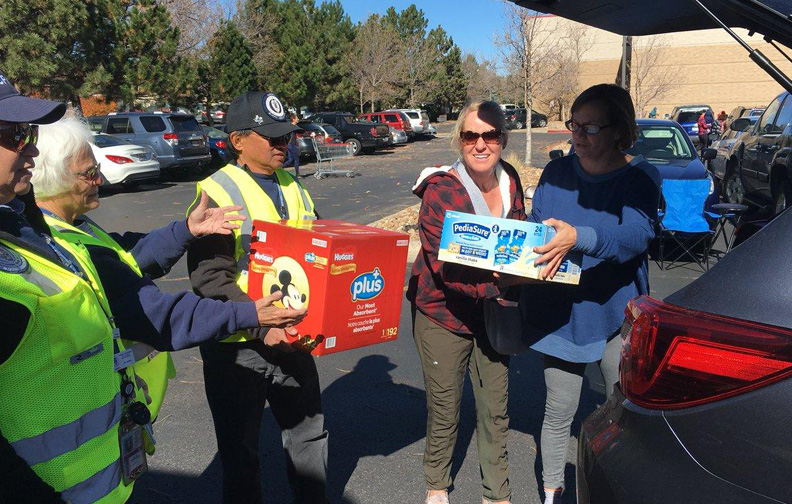}
			&
			\includegraphics[width=0.28\textwidth]{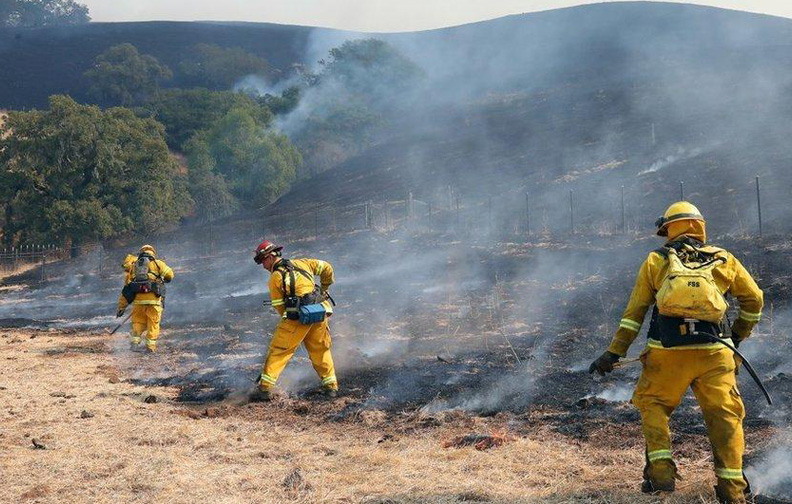}
			& 
			\includegraphics[width=0.28\textwidth]{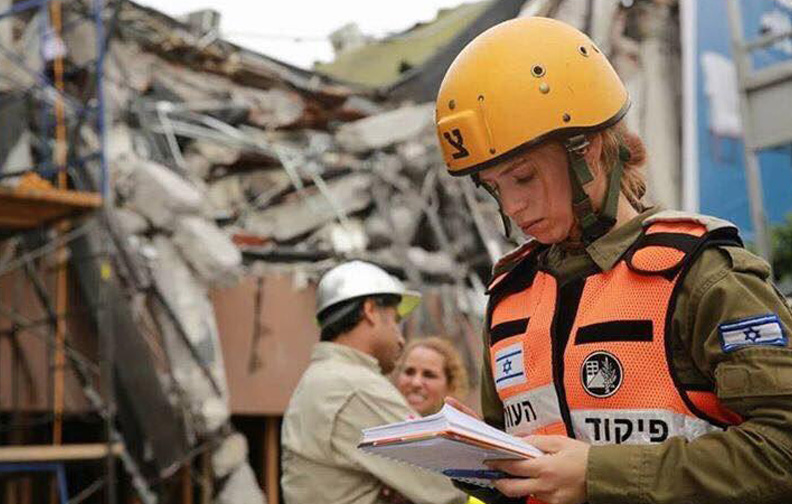}
			\\
			&
			{\scriptsize \textbf{(g)} Puerto Rico donation drive going on until 4 p.m. today and again on Oct. 28! https://t.co/zXZBrHeLCQ https://t.co/2T9k2mTCIs}
			&
			{\scriptsize \textbf{(h)} Raining Ash and No Rest: Firefighters Struggle to Contain California Wildfires https://t.co/G6pkvO53lJ \#SocialMedia https://t.co/DRUCJ7t6G6}
			&
			{\scriptsize \textbf{(i)} Israeli aid team in \#Mexico working day \& night to find survivors \#MexicoEarthquake https://t.co/UO2ZKkaisB }
			\\
			\raisebox{5.8\normalbaselineskip}[0pt][0pt]{\rotatebox[origin=c]{90}{\textbf{Affected individuals}}}
			&
			\includegraphics[width=0.28\textwidth]{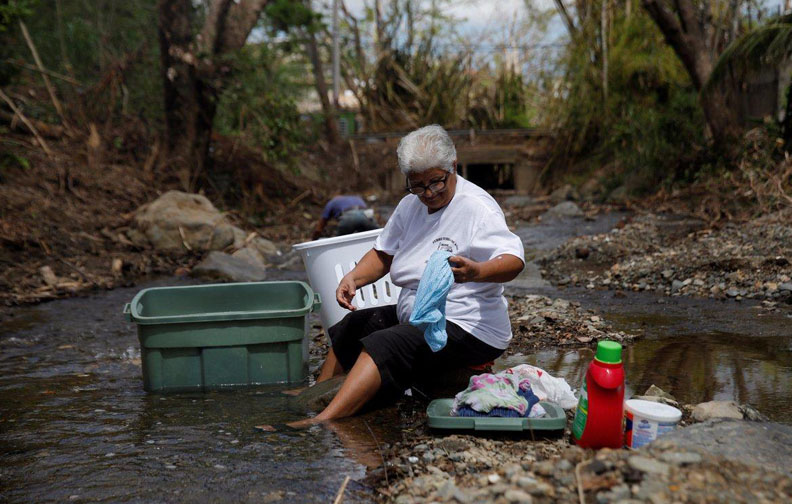}
			&
			\includegraphics[width=0.28\textwidth]{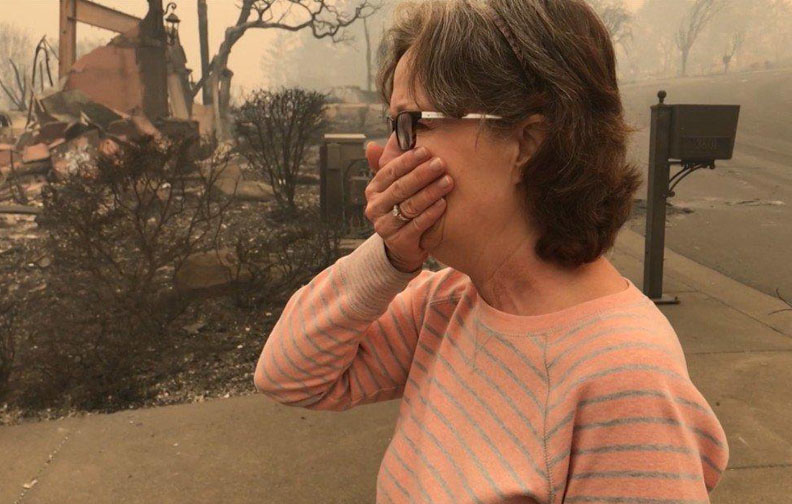}
			&
			\includegraphics[width=0.28\textwidth]{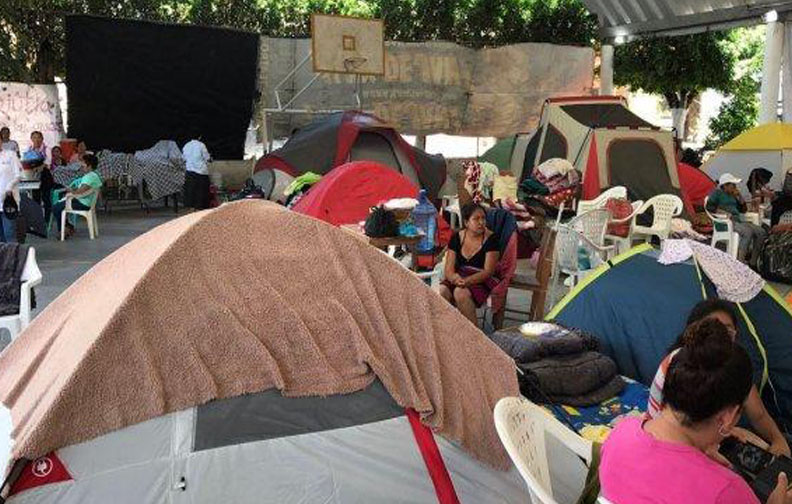}
			\\
			&
			{\scriptsize \textbf{(j)} RT {@}ajplus: 85\% of Puerto Rico remains without power. 40\% of people still don’t have access to drinking water. https://t.co/LKbGc7DI2R}
			&
			{\scriptsize \textbf{(k)} RT @USRealityCheck: Homeowners cry as they return after fire https://t.co/kQIuhBCMQn \#USNews \#USRC https://t.co/A9ozlh2Mx1}
			&
			{\scriptsize \textbf{(l)} In Jojutla, Mexico, earthquake left hundreds homeless and hungry \#TODAY https://t.co/jg6RFv8oHs https://t.co/iHUOYb0eEE}
			\\
			\raisebox{5.4\normalbaselineskip}[0pt][0pt]{\rotatebox[origin=c]{90}{\textbf{Other relevant info}}}
			&
			\includegraphics[width=0.28\textwidth]{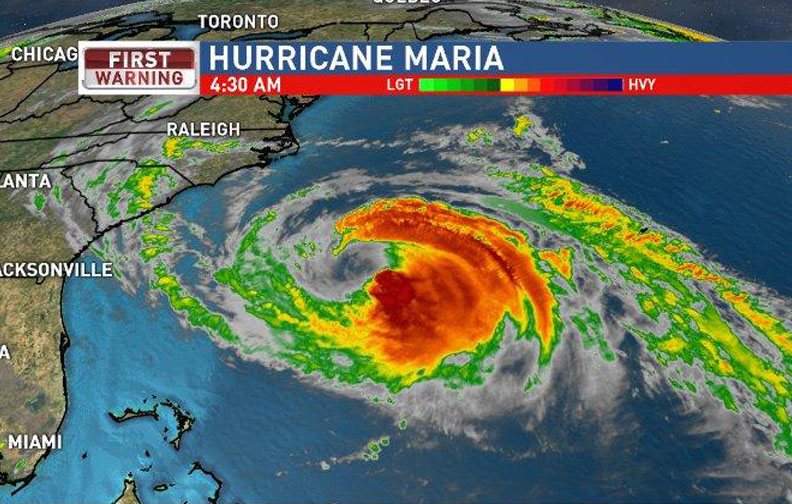}
			&
			\includegraphics[width=0.28\textwidth]{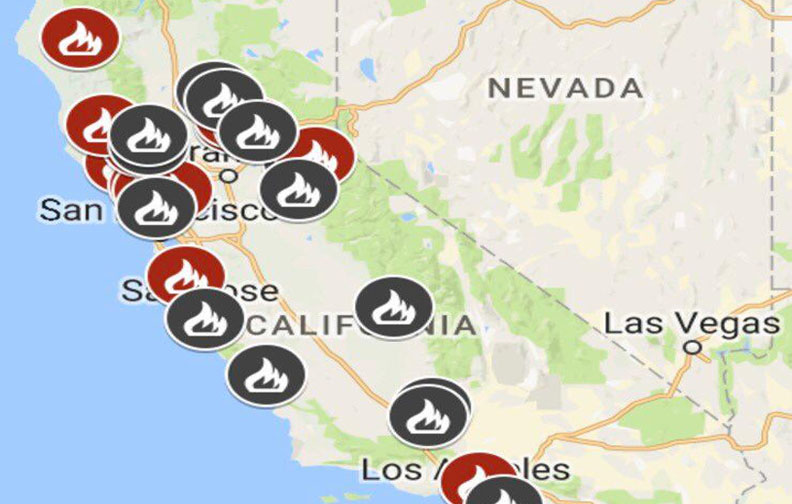}
			&
			\includegraphics[width=0.28\textwidth]{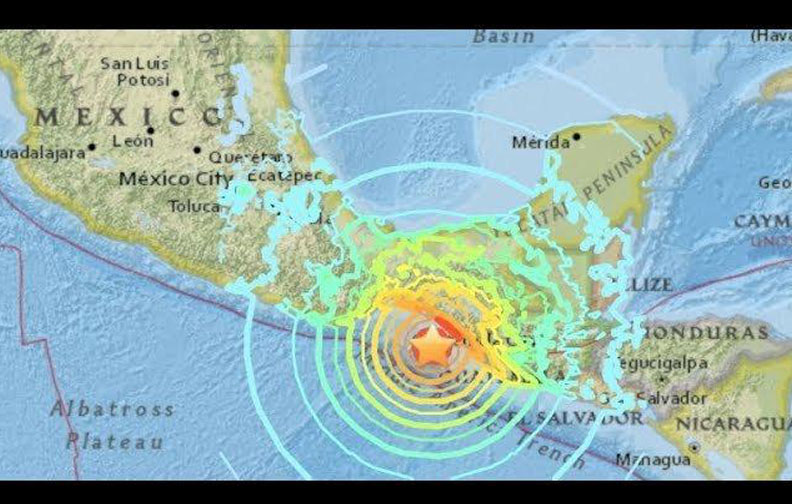}
			\\
			&
			{\scriptsize \textbf{(m)} \#Maria remains a Category 1 Hurricane... Heavy rain by mid-week in the Outer banks https://t.co/Vm4qRPBMkY}
			&
			{\scriptsize \textbf{(n)} California is on fire! Please be safe out there everyone! https://t.co/dnuLv5FayS}
			&
			{\scriptsize \textbf{(o)} Sun-Earthquake Model Matches M8.1 in Mexico https://t.co/GEzzk9tECr https://t.co/48WWjCWw5p}
			\\
			\raisebox{5.2\normalbaselineskip}[0pt][0pt]{\rotatebox[origin=c]{90}{\textbf{Severe damage}}}
			&
			\includegraphics[width=0.28\textwidth]{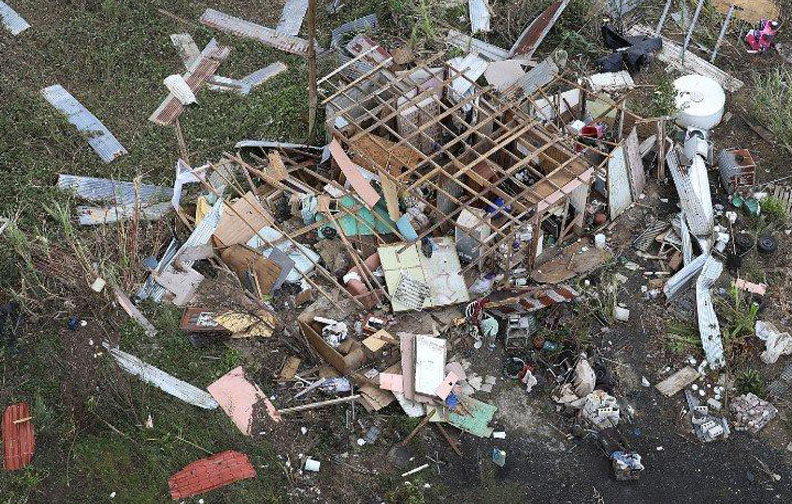}
			&
			\includegraphics[width=0.28\textwidth]{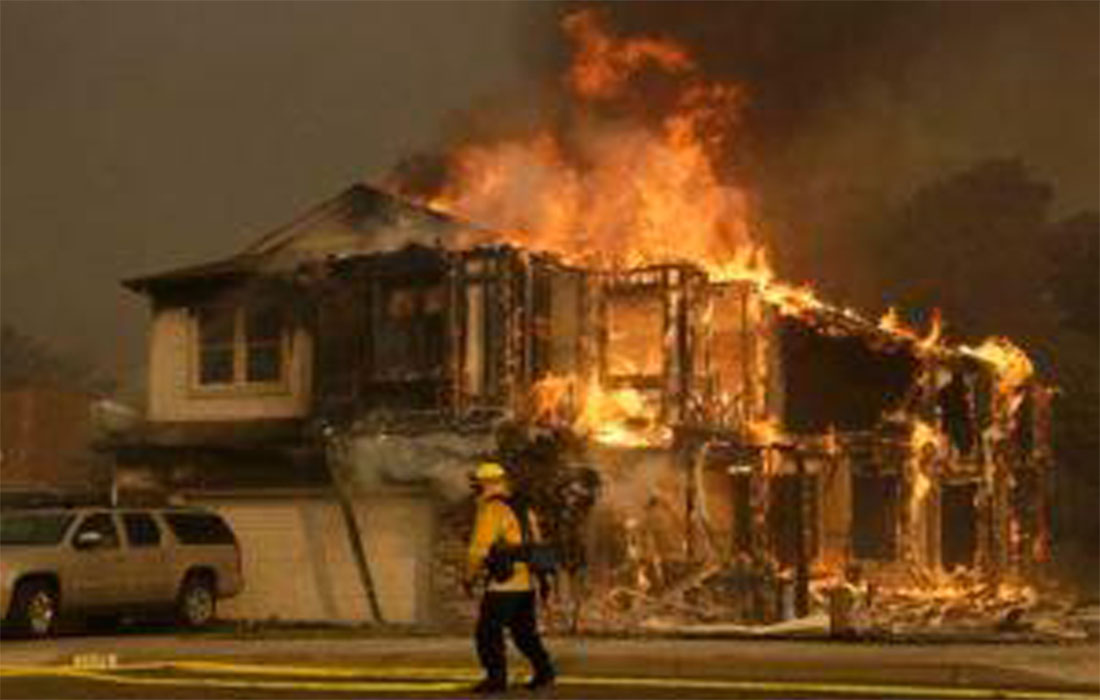}
			&
			\includegraphics[width=0.28\textwidth]{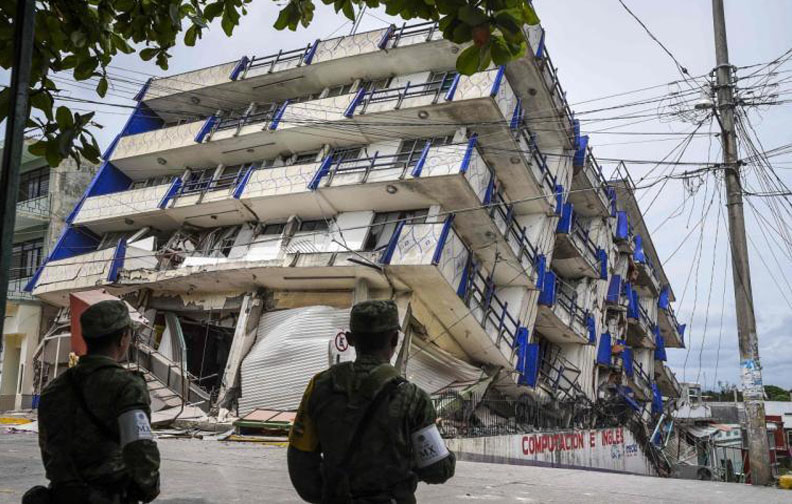}
			\\
			&
			{\scriptsize \textbf{(p)} Corporate donations for Hurricane Maria relief top \$24 million https://t.co/w34ZZziu88 https://t.co/ePddksfFc2}
			&
			{\scriptsize \textbf{(q)} California Wildfires Threaten Significant Losses for P/C Insurers, Moodya Says https://t.co/ELUaTkYbzZ https://t.co/Os8UAAjxGb}
			&
			{\scriptsize \textbf{(r)} Southern Mexico rocked by 6.1-magnitude earthquake CLICK BELOW FOR FULL STORY... https://t.co/Vkz6fNVe5s... https://t.co/Cn4LSWrN4T}
			\\
		\end{tabular}
	}
	\caption{Example tweet text and image pairs with different annotation labels from different disaster events.}
	\label{fig:sample_crisis_images}
	%\vspace{-.2in}
\end{figure*}

\subsection{Crowdsourcing Results and Discussion}
\begin{figure}
	\centering
	\includegraphics[width=1\linewidth]{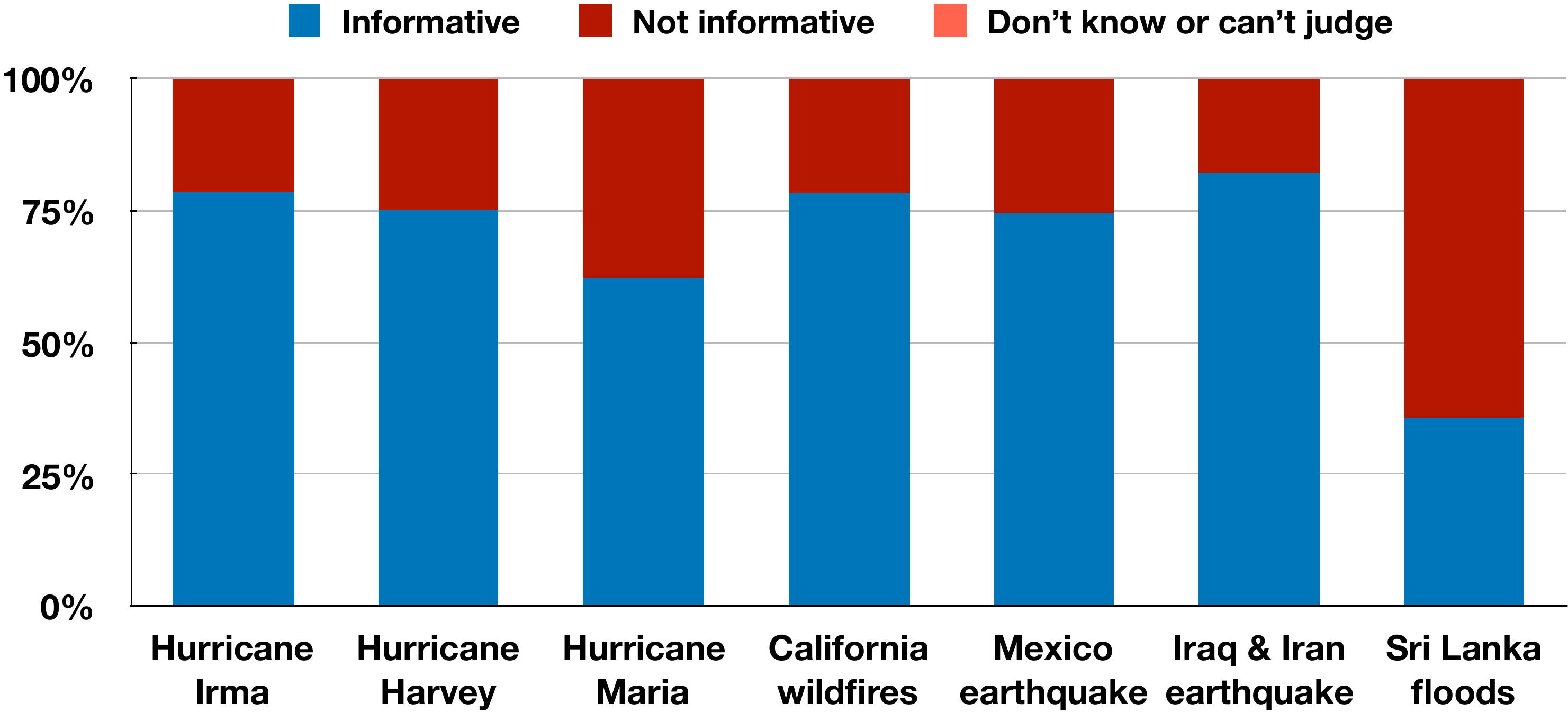}
	\caption{Manual annotation results of tweets (text) for the informative task.}
	\label{fig:info_tweets_dist}
\end{figure}

\begin{figure}
	\centering
	\includegraphics[width=1\linewidth]{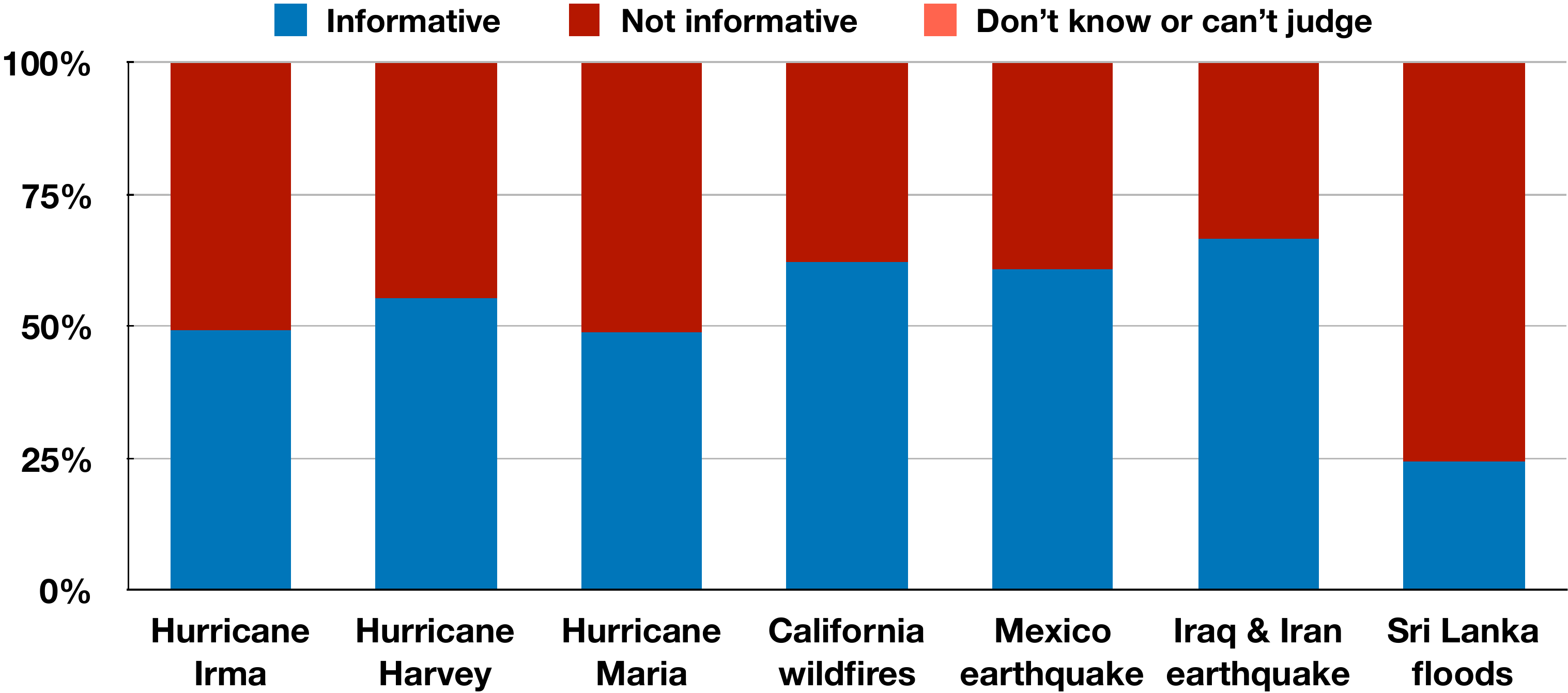}
	\caption{Manual annotation results of images for the informative task.}
	\label{fig:info_image_dist}
\end{figure}

\begin{figure}
	\centering
	\includegraphics[width=1\linewidth]{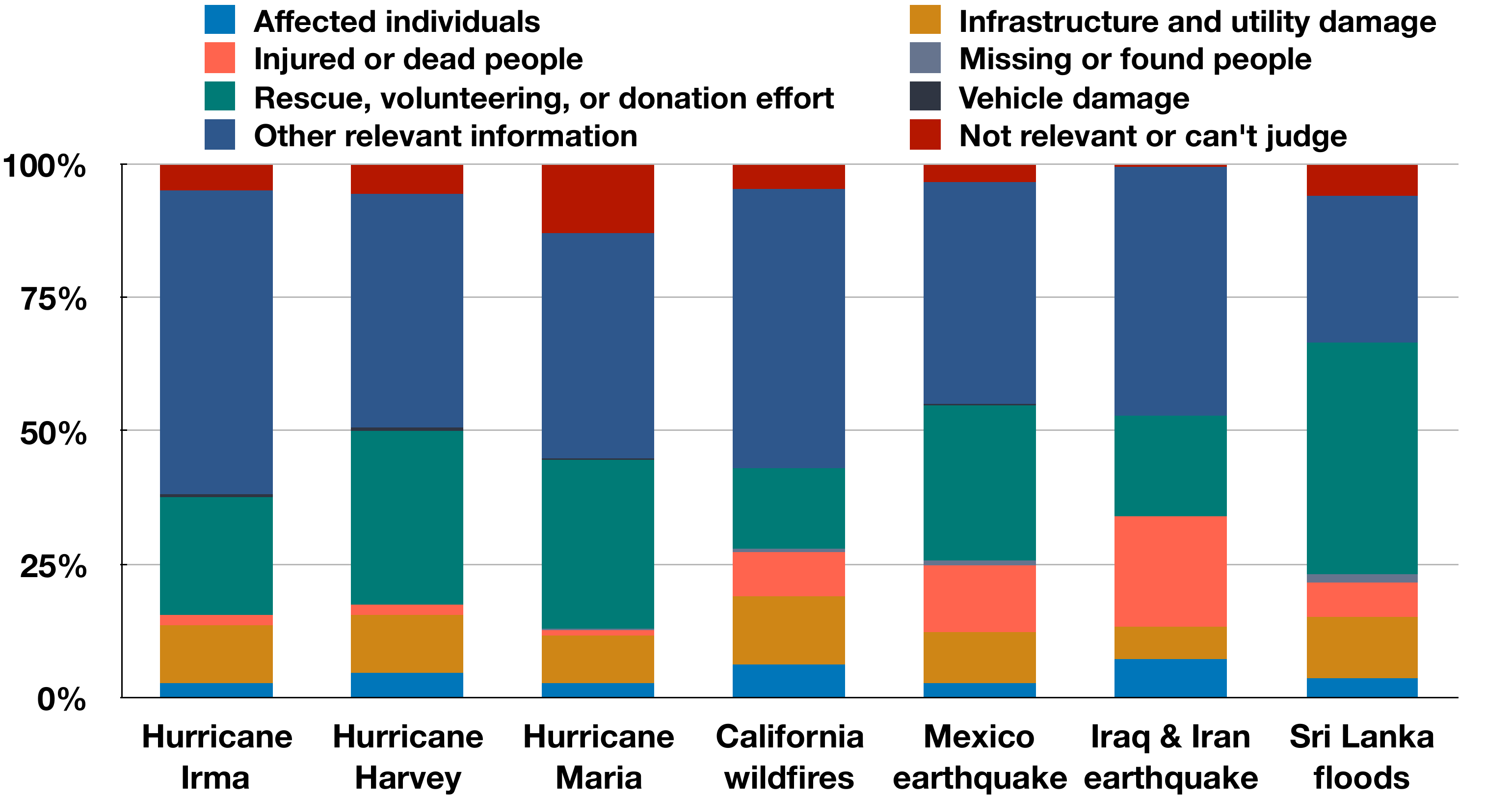}
	\caption{Manual annotation results of tweets (text) for the humanitarian task.}
	\label{fig:hum_tweets_dist}
\end{figure}

\begin{figure}
	\centering
	\includegraphics[width=1\linewidth]{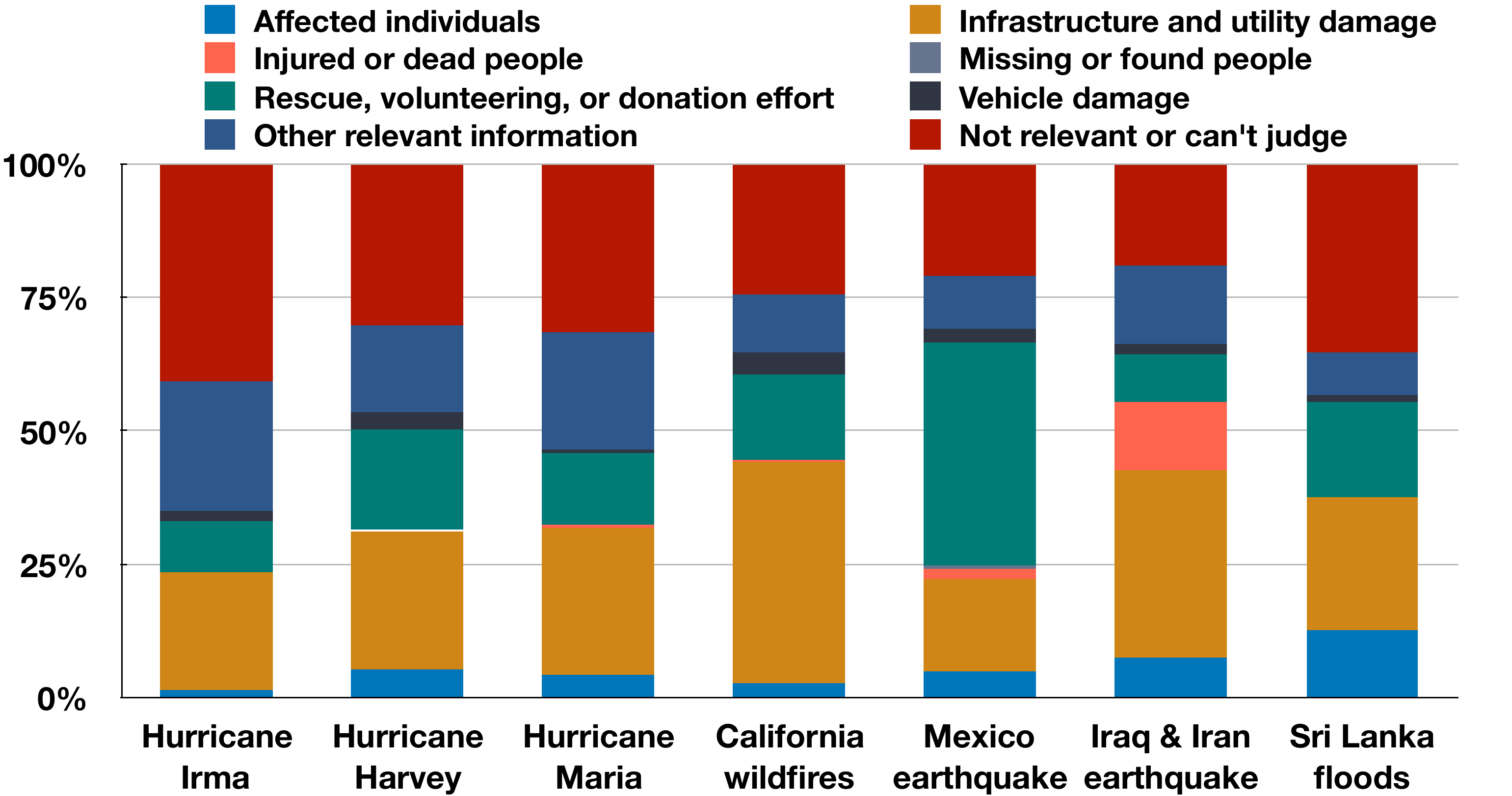}
	\caption{Manual annotation results of images for the humanitarian task.}
	\label{fig:hum_image_dist}
\end{figure}

\begin{figure}
	\centering
	\includegraphics[width=1\linewidth]{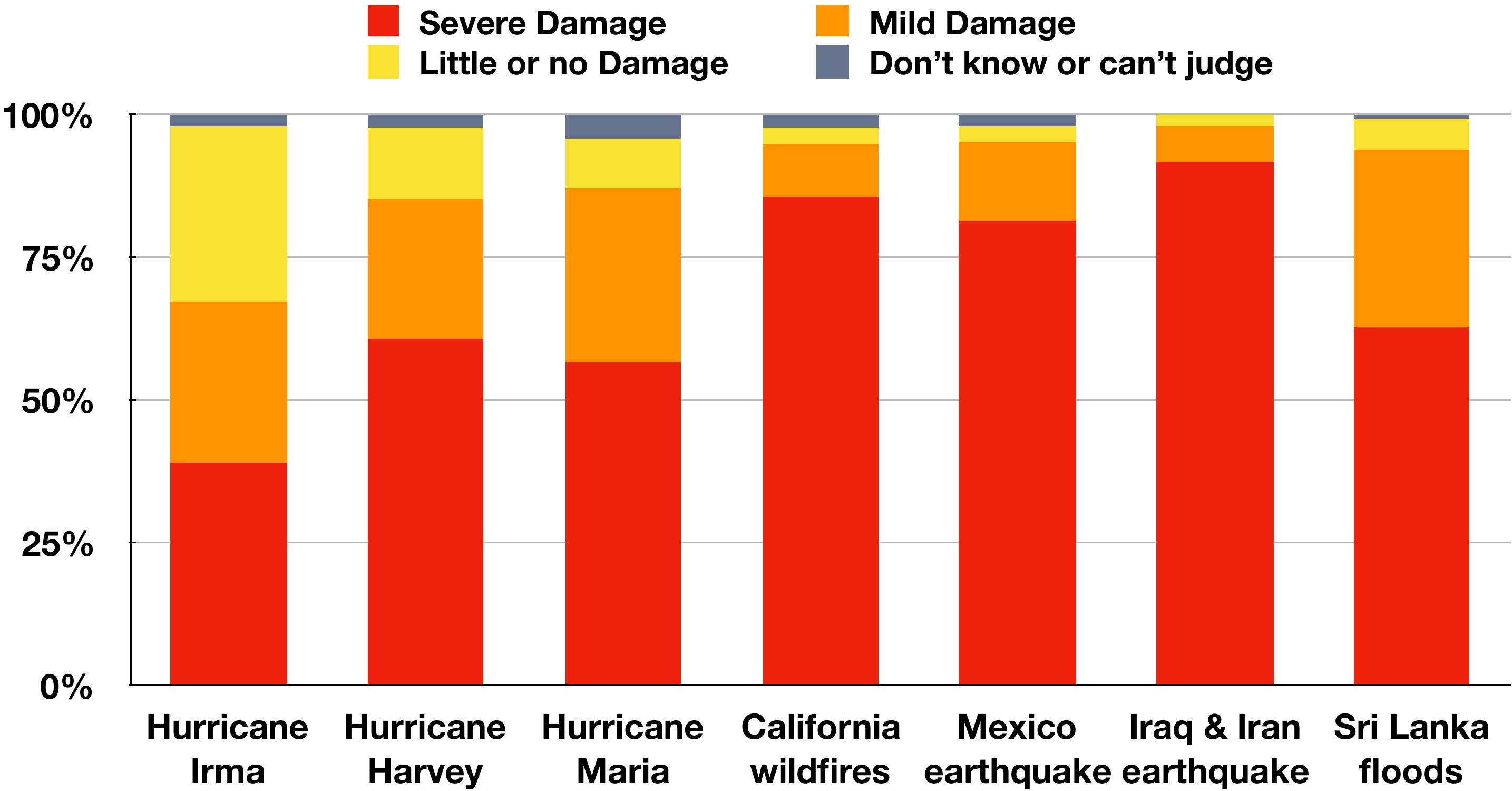}
	\caption{Manual annotation results of images for the damage severity assessment task. %\textcolor{red}{Also the legend needs to be updated to dont know or cant judge instead of not relevant or dont know.}
	}
	\label{fig:damage_image_dist}
\end{figure}

Figure~\ref{fig:sample_crisis_images} illustrates example tweet text and image pairs with different annotations 
from different disaster events. Figures~\ref{fig:info_tweets_dist} and~\ref{fig:info_image_dist} show the distribution of tweet text and image results into the informative categories task, respectively. Similarly, Figures~\ref{fig:hum_tweets_dist} and~\ref{fig:hum_image_dist} show the results of tweet text and image annotations for the humanitarian categories task, respectively. Lastly, Figure~\ref{fig:damage_image_dist} shows the manual annotation results for the damage severity assessment task.

\begin{figure*}[h]
	\centering
	\includegraphics[width=1\linewidth]{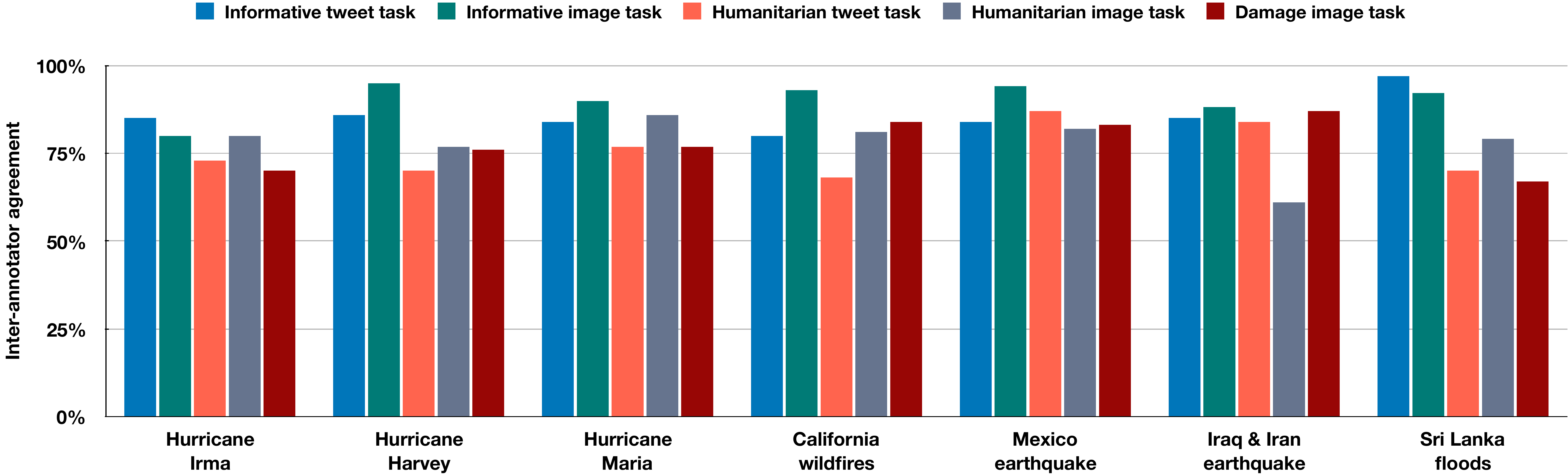}
	\caption{Inter-annotator-agreement scores for all the tasks and all the disaster events.}
	\label{fig:iaa_scores}
\end{figure*}

From around 25\% to 35\% of the tweet text data of all the events is considered as ``not informative'' with an exception of the Sri Lanka floods event in which case the ``not informative'' category is around 60\% (see Figure~\ref{fig:info_tweets_dist}). This finding is in line with previous studies that analyze Twitter data during disasters.
%The problem of finding informative tweets during disasters is akin to finding a needle in a giant haystack.
The prevalence of the ``not informative'' category in the image informative task is higher than the text informative task (see Figure~\ref{fig:info_image_dist}). All informative tweets and images were selected for the second task (humanitarian categories) which we examine next.

In the humanitarian task for tweet text annotations, the ``Other relevant information'' and ``Rescue, volunteering, or donation effort'' categories appear as the most prevalent ones among others. The ``Missing or found people'' category is one of the rarest, as can be seen in Figure~\ref{fig:hum_tweets_dist}. It appears that earthquakes cause more injuries and deaths compared to hurricanes and floods (see Figure~\ref{fig:hum_tweets_dist}). In particular, the reports of ``Injured or dead people" both in the tweets and images in the Iraq-Iran earthquake event are significantly more than any other event.
%This is understandable as both countries (Iraq \& Iran) are underdeveloped compared to the other countries in our dataset.
A small proportion of the ``Not relevant or can't judge'' tweets can still be seen in Figure~\ref{fig:hum_tweets_dist}. This is mainly due to our sampling strategy for this annotation task (i.e., if either text or image of a tweet is informative, it is selected to be annotated for the humanitarian task). 

We observe that images in tweets tend to contain more damage-related information compared to their corresponding text. For instance, according to Figure~\ref{fig:hum_image_dist}, the ``Infrastructure and utility damage'' category is generally prevalent in all the events, however, in the case of the California wildfires, it appears to be around 50\% of the informative event data. Moreover, the ``Vehicle damage" category, which does not appear at all in the text annotation results, appears in the image annotations of many events (see California wildfires and Hurricane Harvey bars in Figure~\ref{fig:hum_image_dist}). The other most prevalent information type present in images is ``Rescue, volunteering, or donation effort''. Mainly, these images show people that help or rescue others, or are involved in volunteering efforts during or after a disaster.

The results of the damage severity assessment task are shown in Figure~\ref{fig:damage_image_dist}. Since we used images that were already annotated as ``Infrastructure and utility damage'', the results do not show many ``Don't know or can't judge'' cases in this task. Most of the images were annotated as severely-damaged infrastructure in all the events. However, most of the severe damage seems to be actually caused by the earthquakes and wildfires as opposed to the hurricanes and floods. Figure~\ref{fig:iaa_scores} shows the inter-annotator-agreement of all the tasks. Generally, all the tasks have strong agreement scores.

\section{Applications and Future Directions}
\label{sec:discussion}
The provided datasets have several potential applications in many different domains. First of all, CrisisMMD datasets can be used in any multimodal task involving computer vision and natural language processing. For instance, one can try to learn a joint embedding space of tweet text and images that can be used for text-to-image as well as image-to-text retrieval tasks. Another multimodal use case of CrisisMMD can be the image captioning task where the goal is to learn a mapping from the visual content to its textual description. Furthermore, more powerful event summarization models can be trained on these aligned and structured multimodal data to automatically generate a multimedia summary of a given event.
%In addition, one can use this dataset when it comes to the problem of text and image based multimodal setting.\textcolor{red}{Previous sentence is not clear...} 
Since we have developed CrisisMMD datasets mainly with the ``humanitarian aid'' use case in mind, we further discuss applications specific to the humanitarian domain in the rest of this section.

\subsection{High-level Situational Awareness by Reducing Information Overload} 
Information posted on social media during natural and man-made disasters vary greatly. Studies have revealed that a big proportion of social media data consists of irrelevant information that is not useful for any kind of relief operations. Humanitarian organizations do not want a deluge of noisy messages that are of a personal nature or those that do not contain any useful information. Instead, they look for messages which contain some useful information. Among other uses, they use the informative messages and images to gain situational awareness. This dataset provides human annotations along informative and not informative messages from seven crisis events to help community to build more robust systems.

\subsection{Critical and Potentially Actionable Information Extraction}
Depending on their roles and mandate, humanitarian organizations differ in terms of their information needs. Several rapid response and relief agencies look for fine-grained information about specific incidents which is also actionable. Such information types include reports of injured or dead people, critical infrastructure damage (e.g., a collapsed bridge), and rescue demand among others. Our dataset provides human annotations along many such critical humanitarian information needs, which can prove to be life saving if more effective systems and computational methods are developed. Furthermore, the damage severity annotations are critical for many response organizations to direct their focus to, for example, severely damaged infrastructure to reduce suffering of affected people. 

Furthermore, with several thousands of manually annotated pairs of tweets and images, we claim that CrisisMMD is the first and largest multimodal dataset to date published for research community to explore different approaches and build computational methods to help humanitarian cause.

%From the seven natural disasters we collected $\sim14$ millions tweets consisting of $\sim57K$ images, which is one of the largest dataset for humanitarian task. From this large collection we have $\sim38K$ manually annotated tweets and associated images. Our annotation consists of three tasks, however, this dataset can also be useful to categorize tweets with crisis type as our data collection covers four crisis types. The system that monitor tweets can train a model for categorizing tweets with crisis types and use them in the subsequent classification processes to provide real-time information. 

%From our study, we realized that the multimodal annotation for social media data is more complex than typical audio-visual modality. It is because in audio-visual scenario, the audio, transcription and visual part are aligned, whereas a user might say about one thing and posts completely different things. Hence, for social media data there is a less alignment between text and image that user post, which makes the task complex. 

% ================
% Conclusions
% ================
\section{Conclusions}
\label{sec:conclusions}
Information available on social media at times of a disaster or an emergency is useful for several humanitarian tasks. Despite extensive research that uses social media textual content, little focus has been given to images shared on social media. One issue in this regard is the lack of labeled imagery data. To address this issue, in this paper, we introduced CrisisMMD, multimodal Twitter corpora consisting of several thousands of manually annotated tweets and images collected during seven major natural disasters including earthquakes, hurricanes, wildfires, and floods that happened in the year 2017 across different parts of the World. 
%The data has been collected in 2017 during seven natural disasters. 
The provided datasets include three types of annotations: informative vs.\ not informative, humanitarian categories, and damage severity categories.
We also presented a number of humanitarian use cases and tasks that can be fulfilled using these datasets if more robust and effective systems are developed. 

%To advance the current state-of-art we made the dataset publicly available. In addition to the labeled dataset, we also plan to release unlabeled dataset that has been collected during these events, which will useful for unsupervised feature learning tasks. Moreover, we also present baseline results focusing only Harvey event. Along with the dataset, we also plan to make the baseline experimental code publicly available, which can help in future studies to design more advanced machine learning model. For the crisis computing our dataset will shade a light in future for multimodal experiments. 

{
\footnotesize
\bibliographystyle{aaai}
\balance
\bibliography{bibliography}
}

\end{document}